# Stabilization of $Mn^{4+}$ in synthetic slags and identification of important slag forming phases

Alena Schnickmann[a,*], Danilo Alencar De Abreu[b], Olga Fabrichnaya[b] and Thomas Schirmer[a]

[a] *Department of Mineralogy, Geochemistry, Salt Deposits, Institute of Disposal Research, Clausthal University of Technology, Adolph-Roemer-Str. 2A, 38678 Clausthal-Zellerfeld, Germany; thomas.schirmer@tu-clausthal.de*

[b] *Institute of Materials Science, TU Bergakademie Freiberg, Gustav-Zeuner Stra. 5, 09599, Freiberg, Germany; Olga.Fabrichnaya@iww.tu-freiberg.de*

*\* Correspondence: alena.schnickmann@tu-clausthal.de; Tel.: +49-5323-722051*

## Abstract

The expected shortage of Li due to the strong increase in electromobility is an important issue for the recovery of Li from spent Li-ion batteries. One approach is pyrometallurgical processing, during which ignoble elements such as Li, Al and Mn enter the slag system. The Engineered Artificial Minerals (EnAM) strategy aims to efficiently recover critical elements. This study focuses on stabilizing Li-manganates in a synthetic slag and investigates the relationship between $Mn^{4+}$ and Mg and Al in relation to phase formation. Therefore, three synthetic slags (Li, Mg, Al, Si, Ca, Mn, O) were synthesized. In addition to $LiMn^{3+}O_2$, $Li_2Mn^{4+}O_3$ was also stabilized. Both phases crystallized in a Ca-silicate-rich matrix. In the structure of $Li_2MnO_3$ and $LiMnO_2$, Li and Mn can substitute each other in certain proportions. As long as a mix of $Mn^{2+}$ and $Mn^{3+}$ is present in the slag, spinels form through the addition of Mg and/or Al.

# 1. Introduction

## 1.1 Recovery of critical elements from the slag phase

The usage of critically relevant elements, such as Li, will be increasingly in demand due to their use in various technological areas [1]. The EU has established a list of critical elements, which includes Li as well as Co [2]. In Li-ion batteries (LIBs), these two elements are used as cathode materials ($LiCoO_2$). The geological supply of such elements is limited, thus the recovery of e.g. Li from the industrial waste stream such as from old LIBs is indispensable [3–5]. Pyrometallurgical processes [6] are a promising method for the recovery. With this route, various elements (e.g. Co, Ni or Cu) can be recovered directly, while base elements (e.g. Li, Mg, Al or Mn) are either lost as dust or enter the slag phase and form complex compounds. For efficient recovery of critical elements such as Li, it is essential to concentrate the element of interest in a single phase within the slag [1,7]. Due to the high $O_2$ affinity of Li, this element should only crystallize in a Li-rich phase. In addition, this phase should have a high Li-content as well as good processing properties (e.g., habitus, crystal size, magnetic properties). The stabilization of Li in a single phase for an efficient recovery of critically relevant elements is the idea behind: Engineered Artificial Minerals (EnAM) [8–10] (Figure 1).

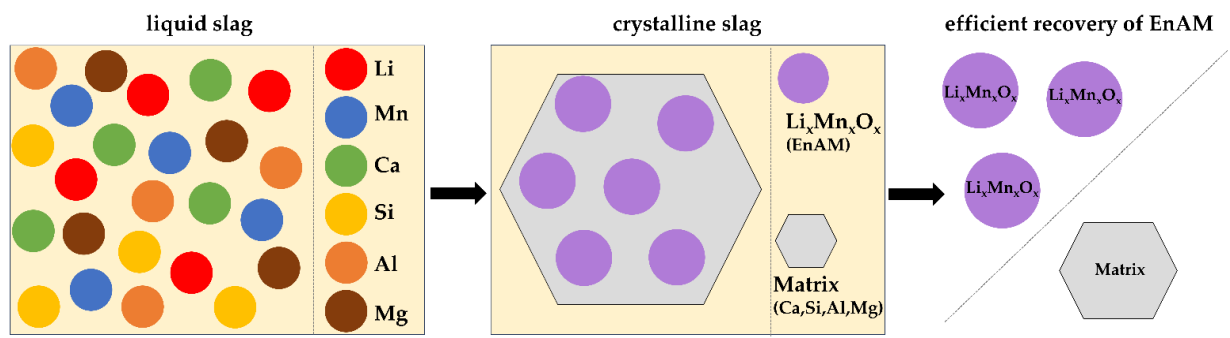

*Figure 1: Sketch of the EnAM strategy for efficient Li recovery. The aim is to form a Li-manganate as an early crystallizate from the liquid melt. The other elements (Mg, Al, Si, Ca) should form matrix-forming phases. This early crystallizate (Li-Manganate) should be efficiently separated from the matrix.*

## 1.2 Potential EnAMs for efficient recovery

Studies by Elwert et al. [11] shown that the Li-aluminate $LiAlO_2$ would be suitable as a potential EnAM, as the idiomorphic to hypidiomorphic crystals can be separated from the remaining slag via flotation [12]. In the presence of especially Mg, Al and Mn, spinel complexes are formed during solidification and solid solutions occurred between spinels (e.g. $MgMn_2O_4$, $MgAl_2O_4$, $MnAl_2O_4$ and $Mn^{2+}Mn^{3+}_2O_4$), which hampered the formation of $LiAlO_2$ [8,13–15]. In addition, this Li aluminate can contain up to 3 wt.% Si, which complicates hydrometallurgical processing [12]. As the use of Mn in new Li batteries increases, Mn will become part of a complex slag system along with Li, Mg and Al. In addition, small amounts of Mg and Al in $LiMnO_2$ and $Li_2MnO_3$ are expected to promote intralayer diffusion and Mg should also promote interlayer diffusion [16]. For this reason, current research is focusing on the stabilization of Li in Li-manganates as a potential EnAM [9]. The difficulties here are the redox-sensitive behavior of Mn and the Jahn-Teller effect on the $Mn^{3+}$ at temperatures above 1445 K [14,15]. Schnickmann et al. [9] showed that it is possible to stabilize a Li-manganate in a complex synthetic oxide slag system (Li, Mg, Al, Si, Ca, Mn). Under normal atmosphere a mixture between $Mn^{2+}$ and $Mn^{3+}$ was present and pure $LiMnO_2$ has formed. Nevertheless, the formation of spinels and spinel solid solutions could not yet be prevented, as well as the incorporation of $Mn^{2+}$ into the Ca-silicate matrix. In a follow-up experiment, pure oxygen (100 % $O_2$) was used to stabilize a higher Mn oxidation stage in the slag. For this experiment, the same batch precursors were used as in Schnickmann et al. [9]. This experiment is designed to investigate whether it is possible to stabilize higher Mn oxidation states in synthetic slags and whether $Mn^{4+}$ forms a compound with the other elements, especially with Mg and/or Al.

1.3 Important slag forming phases

In a complex slag system consisting of $Li_2O$, MgO, $Al_2O_3$, $SiO_2$, CaO, and MnO, mainly binary and ternary oxide compounds are formed, followed by a residual melt [8,9]. The compounds described below represent possible phases in a crystalline slag with a special focus on Li-rich compounds and the Mn oxides. Compounds between $Mn^{4+}$ and Al/Mg have not yet been described in the literature. The ternary compounds are of minor importance for the results presented and are therefore not discussed in detail in this chapter.

The $Li_2O$-$MnO_x$ system ($1 \leq x \leq 2$) must be considered for the investigation of possible Mn-containing EnAMs. Due to the redox sensitivity of Mn, the formation of various Li-manganates with different concentrations and Mn speciations are possible. The phase equilibria in $Li_2O-MnO_x$ system at air is shown in Figure 2. According to the experimental data obtained by Paulsen and Dahn [17], cubic spinel transforms to tetragonal spinel at high temperatures. A miscibility gap between hausmannite ($Mn_3O_4$) and *t*-spinel was proposed. $LiMnO_2$ is stable at temperatures higher than 1223 K, while the $Li_2MnO_3$ phase is stable up to 1240 K, approximately. In addition to Paulsen and Dahn [17], Longo et al. [18] have also described such compounds in detail, as well as the influence of temperature, pressure and pH on the formation and stability of such Li-manganates. Below 400 °C, the cubic spinel phases with stoichiometry $LiMn_{1.75}O_4$ and $Li_4Mn_5O_{12}$ are found stable. They can also be part of the spinel solid solution, as shown in the calculations of [19] but to achieve equilibrium at these low temperatures would be problematic. The phase $Li_2Mn_3O_7$ is calculated to be stable, but its stability has not yet been experimentally proven [17,18,20]. Between 400 and 800 °C, the Li-spinel ($Li_{(1+x)}Mn_{(2-x)}O_4$) has a large stability field between $LiMn_2O_4$ ($Mn^{3.5+}$) and $Li_4Mn_5O_{12}$ ($Mn^{4+}$) and can coexist with $Li_2MnO_3$ ($Mn^{4+}$). Above 800 °C, $Li_2MnO_3$ and $LiMn_2O_4$ decompose into $Mn_3O_4$ and $O_2$ alongside layered $LiMnO_2$. From 1000 to 1060 °C, orthorhombic $LiMnO_2$ can coexist with lithiated hausmannite ($Li_xMn_{3-x}O_4$) [17]. According to Mishra and Ceder [21], the Jahn-Teller distortion affects the tetragonal spinel $LiMn_2O_4$ as well as orthorhombic (e.g., $LiMnO_2$) and monoclinic layered structures [21]. In addition, the investigations by Schnickmann et al. [9] showed that in the $LiMnO_2$ system the Li and Mn contents can be exchanged and that this compound can incorporate up to 0.35 wt.% Al.

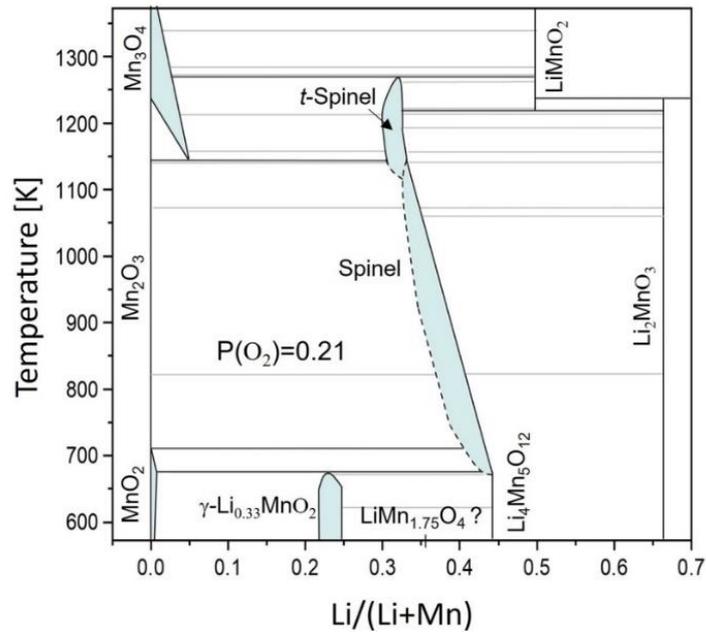

*Figure 2: Adapted phase diagram of the Li$_2$O-MnO$_x$ system at air based on experimental data obtained from Paulsen and Dahn [17].*

Due to their good flotation properties, Li-aluminates have been investigated as potential EnAMs for a long time. Recently thermodynamic assessments of the Li$_2$O-Al$_2$O$_3$ system using CALPHAD approach have been published by Konar et al. [22] and De Abreu et al. [23]. Moreover, De Abreu et al. experimentally studied the phase equilibria in the Al$_2$O$_3$ rich part of the diagram and made calorimetric measurements of heat capacity for intermediate phases. The calculated phase diagram is shown in Figure 3a. Three stable intermediate phases were proposed: LiAl$_5$O$_8$, LiAlO$_2$ and Li$_5$AlO$_4$. LiAlO$_2$ has already been extensively analyzed and discussed for its potential as an EnAM [8,22–24]. Two polymorphs are reported for LiAlO$_2$: low temperature α-phase (trigonal) and the high temperature γ-phase (tetragonal). Two stable modifications of LiAl$_5$O$_8$ with spinel structure but with different space groups were found in the alumina-rich side of the phase diagram. High temperature modification of spinel is inversed and stable with some homogeneity range. No evidences of LiAl$_{11}$O$_{17}$ were found in the microstructures and thus, LiAl$_{11}$O$_{17}$ is not treated as stable.

Another important area of Li-rich phases is the Li$_2$O-SiO$_2$ system, with the following stable phases: Li$_2$Si$_2$O$_5$, Li$_2$SiO$_3$, Li$_4$SiO$_4$ and Li$_8$SiO$_6$ (ordered by decreasing Si concentration). According to the latest results, Li$_6$Si$_2$O$_7$ phase is treated as metastable. Li$_2$SiO$_3$ is formed from a Li$_2$O-SiO$_2$ rich melt. Studies by Chakrabarty et al. [25] shown that at different cooling rates an initial dominance of the thermodynamic driving force occurs, followed by kinetic forces.

Spinels are the most important Li-free compounds, whereas hausmannite ($Mn^{2+}Mn^{3+}_2O_4$) is the most important compound in the $MnO-Mn_2O_3$ system. Below 1172 °C, the Jahn-Teller effect has an influence on the $Mn^{3+}$ position and causes deformation of the crystal parameters. The transformation from tetragonal to cubic is reversible and crystal lattice changes from cubic to tetragonal on cooling. Hausmannite can be regarded as a low temperature phase as well as a deformed spinel under normal conditions (1 atm; 25 °C) and low oxygen partial pressure. Hausmannite forms from $Mn_2O_3$ at 1445 K and air oxygen partial pressure. It can form solid solutions with other spinel compounds. One of the most important spinel in the $MnO-Al_2O_3$ system is galaxite ($MnAl_2O_4$) [14,15,22]. The calculated phase diagrams with thermodynamic parameters optimized by [15] for the $Al_2O_3-MnO_x$ system at air and in presence of metallic Mn are shown in Figure 3b/c, respectively. In the presence of metallic Mn, thus varying oxygen partial pressure $p(O_2)$ with temperature, homogeneity range of the spinel phase is narrow and its stoichiometry is close to $MnAl_2O_4$. According to crystallographic data the spinel phase is normal at low temperature with tetrahedral sites occupied by $Mn^{2+}$ cations, but with temperature increase degree of inversion (fraction of $Al^{+3}$ in tetrahedral site) increases. According to calculations, which agree with experimental results, the homogeneity range of spinel extends with increase of oxygen partial pressure and at air condition ($p(O_2) = 0.21$ bar) homogeneity range of cubic spinel extends from $Mn_3O_4$ to $MnAl_2O_4$ (at high temperature). It should be noted that $Mn_3O_4$ spinel forming from $Mn_2O_3$ by oxygen release has tetragonal structure. The tetrahedral sites are occupied by $Mn^{+2}$ and octahedral by $Mn^{+3}$. Tetragonal spinel transforms to cubic at 1443 K probably due to disproportionation reaction $Mn^{+3} \rightarrow Mn^{+2} + Mn^{+4}$ to reduce Jahn-Teller distortion caused by $Mn^{+3}$.

Phase diagrams of other important systems containing the spinel phase, like $MgO-MnO_x$ and $MgO-Al_2O_3$ systems are also shown in Figure 3d/e, respectively. Thermodynamic parameters obtained by [26] and [27] were applied to calculate the phase relationships for these binary systems. It should be noted that spinel with tetragonal and cubic structures were found in the $MgO-MnO_x$ system. Tetrahedral spinel has homogeneity range from $Mn_3O_4$ to $MgMn_2O_4$, while cubic spinel extends from $Mn_3O_4$ to $Mg_2MnO_4$.

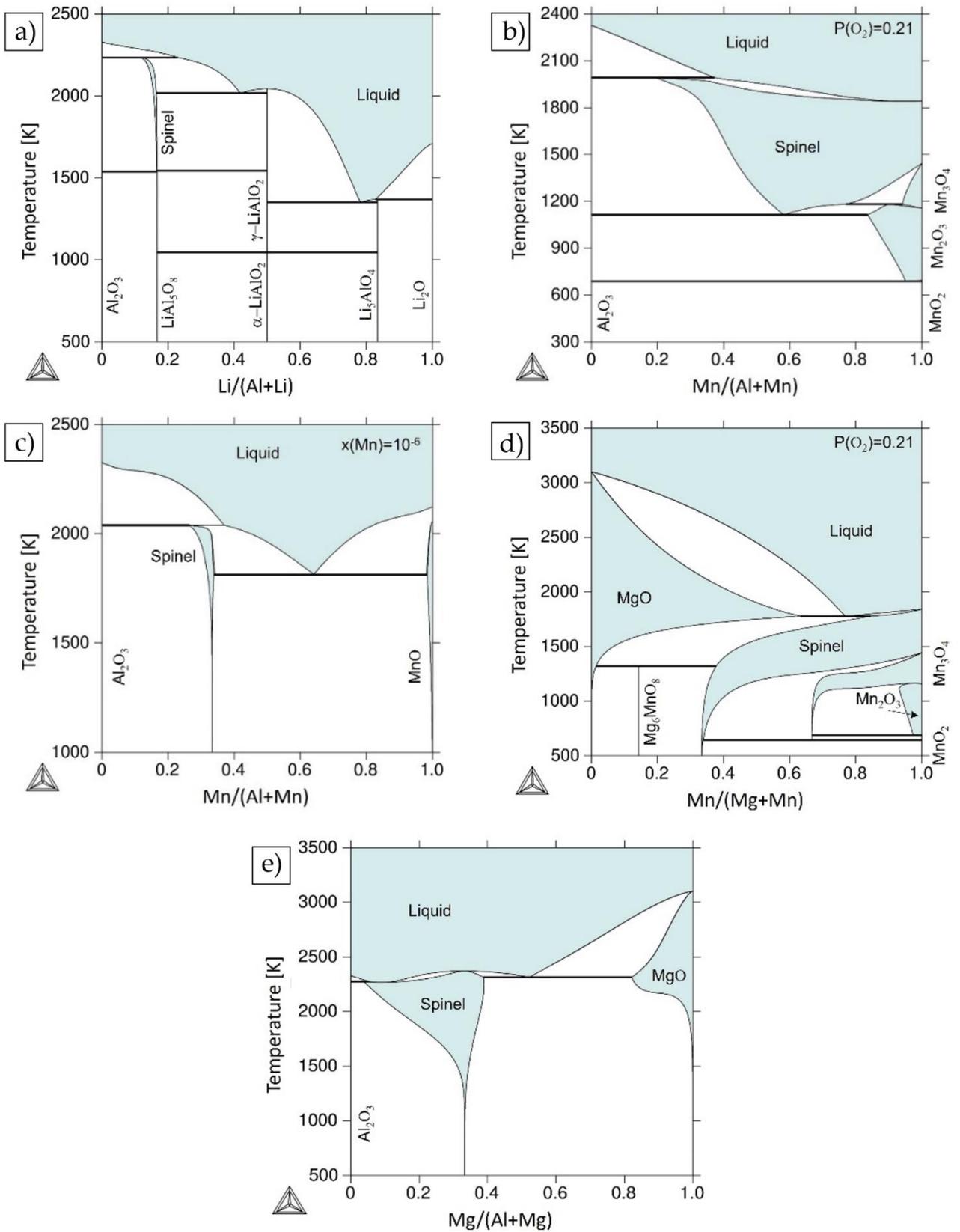

*Figure 3: Calculated phase diagrams of the: a) $Li_2O$-$Al_2O_3$ system, b) $Al_2O_3$-$MnO_x$ system at air; c) $Al_2O_3$-$MnO_x$ system in presence of metallic Mn; d) MgO-$MnO_x$ system at air and (e) $Al_2O_3$-MgO system with thermodynamic parameters optimized from [15,23,26,27].*

## 2. Material and methods

### 2.1 Chemicals and preparation of precursors and reference materials

For the melting experiment, the same batch of precursors was used as in the study by Schnickmann et al. [9], with the following chemical compositions.: $LiNO_3$ (99.5%, Roth, Karlsruhe, Germany), $Mg(NO_3)_2 * 6\ H_2O$ (98%, Carl Roth, Karlsruhe, Germany), $Al(NO_3)_3 * 9\ H_2O$ (98%, Carl Roth, Karlsruhe, Germany), Köstrosol® 0830 AS (Chemiewerk Bad Köstritz GmbH, Bad Köstritz, Germany), $Ca(NO_3)_2 * 4\ H_2O$ (99%, Carl Roth, Karlsruhe, Germany), $Mn(NO_3)_2 * 4\ H_2O$ (98%, Carl Roth, Karlsruhe, Germany) and $C_6H_8O_7$ (99.5%, VWR chemicals, Darmstadt).

The gel-combustion method was used to prepare the precursors. For this purpose, common procedures [28,29] were adapted to the experimental setup in the laboratory. The method has also been described by Schnickmann et al. [9]. The gel-combustion product was thermally treated with $NH_4NO_3$ in quartz crucibles up to 480 °C (10 °K/min; Nabertherm LE 1/11/R7, Nabertherm GmbH, Lilienthal, Germany) to remove the residual carbon. In Table 1 the elemental composition of the precursors can be found.

*Table 1: Elemental composition of the precursors in wt.% and at.%. The precursors have the same composition as [9].*

|         | Precursor 1 | | Precursor 2 | | Precursor 3 | |
|---------|------|------|------|------|------|------|
| Element | wt.% | at.% | wt.% | at.% | wt.% | at.% |
| Li      | 1.9  | 6.3  | 1.9  | 6.3  | 1.8  | 5.8  |
| Mg      | 0.0  | 0.0  | 1.2  | 1.1  | 1.2  | 1.1  |
| Al      | 0.0  | 0.0  | 0.0  | 0.0  | 1.1  | 0.9  |
| Si      | 19.3 | 15.9 | 19.1 | 15.6 | 18.9 | 15.4 |
| Ca      | 29.9 | 17.2 | 29.4 | 16.9 | 29.0 | 16.7 |
| Mn      | 9.8  | 4.1  | 9.2  | 3.9  | 8.6  | 3.7  |
| O       | 39.1 | 56.4 | 39.2 | 56.2 | 39.4 | 56.4 |
| total   | 100.0| 100.0| 100.0| 100.0| 100.0| 100.0|

### 2.2 Melting program

For the melting experiment, the thermally treated precursors were melted in Pt/Rh crucibles under pure oxygen atmosphere at 200 l per hour in a chamber furnace (Nabertherm HT16/17, Nabertherm GmbH, Lilienthal, Germany). The melt program of Wittkowski et al. [8], Schirmer et al. [12] and Schnickmann et al. [9] was adapted for this process (Figure 4).

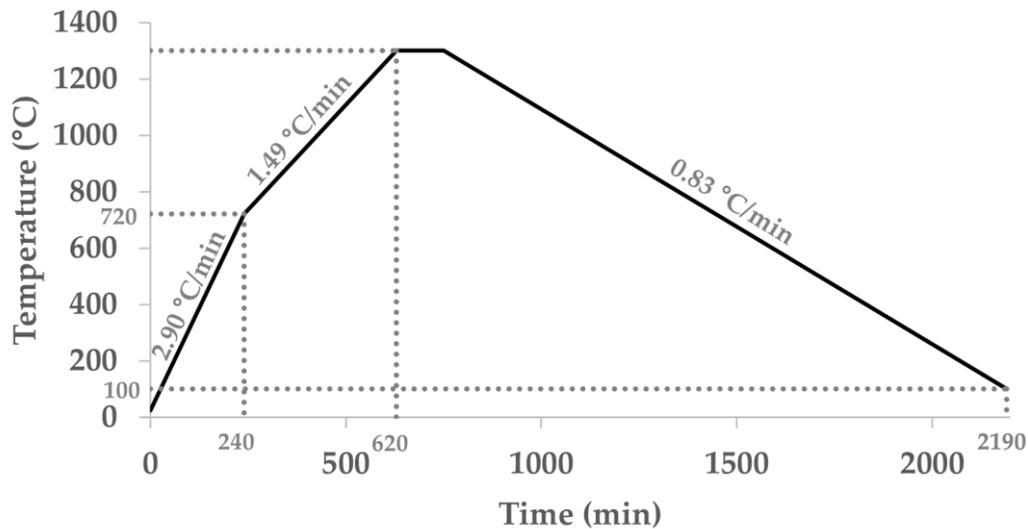

*Figure 4: used heating program for the experiments. Slow heating of the precursor (2.90 °C/min; 1.49 °C/min) should minimize the loss of Li. By cooling down more slowly, the phases should have enough time to form crystals that are as large as possible. During the entire melting experiment, the oxygen supply was 200 l/h.*

2.3 Characterization techniques

The mineralogical composition of the slags was quantitatively and qualitatively identified using powder X-ray diffraction (PXRD and electron probe microanalysis (EPMA).

The crystalline phases in the slag sample were analyzed with PXRD (Panalytical X-Pert Pro Diffractometer). For the measurement (2θ-angle range from 5 ° to 100 °; step size 0.0066 °; 150.45 seconds per step) a Co-X-ray tube (Malvern Panalytical GmbH, Kassel, Germany, λ = 1.7902 Å, 40 kV, 40 mA) has been used. The individual crystalline phases were classified via the American Mineralogist Crystal Structure Database [30] and pdf-2 ICCD XRD database [31].

The elemental composition of single grains/crystals in the prepared thin sections were determined with EPMA (Cameca SXFIVE FE Field Emission, CAMECA SAS, Gennevilliers Cedex, France) using the Kα lines (Mg, Al, Si, Ca, Mn). For the measurement (15 kV beam diameter 100 - 600 nm; Schottky type [32]), the device was calibrated beforehand with certified reference materials (CRM: P&H Developments Ltd; Glossop, Derbyshire, UK and Astimex Standards Ltd; Toronto, ON, Canada). The measured intensities of the emitted X-rays were evaluated using the X-PHI model [33]. With the chosen method, the Li content cannot be analyzed quantitatively with the required precision. Therefore the element concentration of Li was determined using virtual compounds according to T. Schirmer [34].

In order to validate the measurement method and obtain consistent results, repeated measurements were carried out, also on different days, using the international standard rhodonite ($MnSiO_3$; Astimex). The low standard deviation (0.03) of Mn indicates that this element content can be analyzed very precisely and is suitable for calculating the Li content via virtual compounds (Table 2).

*Table 2: The CRM rhodonite was used to verify the measurement accuracy of Mn. The Mn content can be determined with an inaccuracy of ± 0.03 %. Indication in wt.%. *The iron content was not measured.*

| wt% | Average Rhodonite | %StDev. Rhodonite | Ref. Rhodonite |
| --- | --- | --- | --- |
| MgO | 1.92 | 0.01 | 2.0 |
| $Al_2O_3$ | 0.011 | 0.002 | n.a. |
| $SiO_2$ | 46.76 | 0.06 | 46.8 |
| CaO | 4.74 | 0.04 | 4.6 |
| MnO | 42.30 | 0.03 | 42.3 |
| FeO* | 4.34 | n.a. | 4.3 |

## 3. Results

### 3.1 PXRD

In all three samples, wollastonite ($CaSiO_3$; ICDD PDF2: 01-084-0654) as well as the Li-manganates $LiMnO_2$ (ICDD PDF2: 00-035-0749) and $Li_2MnO_3$ (ICDD PDF2: 01-081-1953) could be detected with PXRD. Rankinite ($Ca_3Si_2O_7$; ICDD PDF2: 01-076-0623) was detected in slag 1 and 2 and larnite ($Ca_2SiO_4$; ICDD PDF2: 00-029-0369) in slag sample 3. The presence of hausmannite ($Mn^{2+}Mn^{3+}_2O_4$; ICDD PDF2 00-024-0734) and the Li-silicate $Li_2SiO_3$ (ICDD PDF2: 00-029-0828) is conceivable. However, the presence of the two phases cannot be clearly verified with this method due to line overlaps. Schnickmann et al. [9] have already discovered that within the Li-manganate $LiMnO_2$ the elements Li and Mn can replace each other, which can result in a shifting to smaller or larger lattice parameters. The same applies to the replacement of $Mn^{2+}$ in Ca-silicate matrix. An exchange of elements in the crystal lattice is only possible as long as the exchangeable elements have approximately the same ionic radii (e.g. [35]). An overview of the recorded diffractograms of the three slags is given in Figure 5.

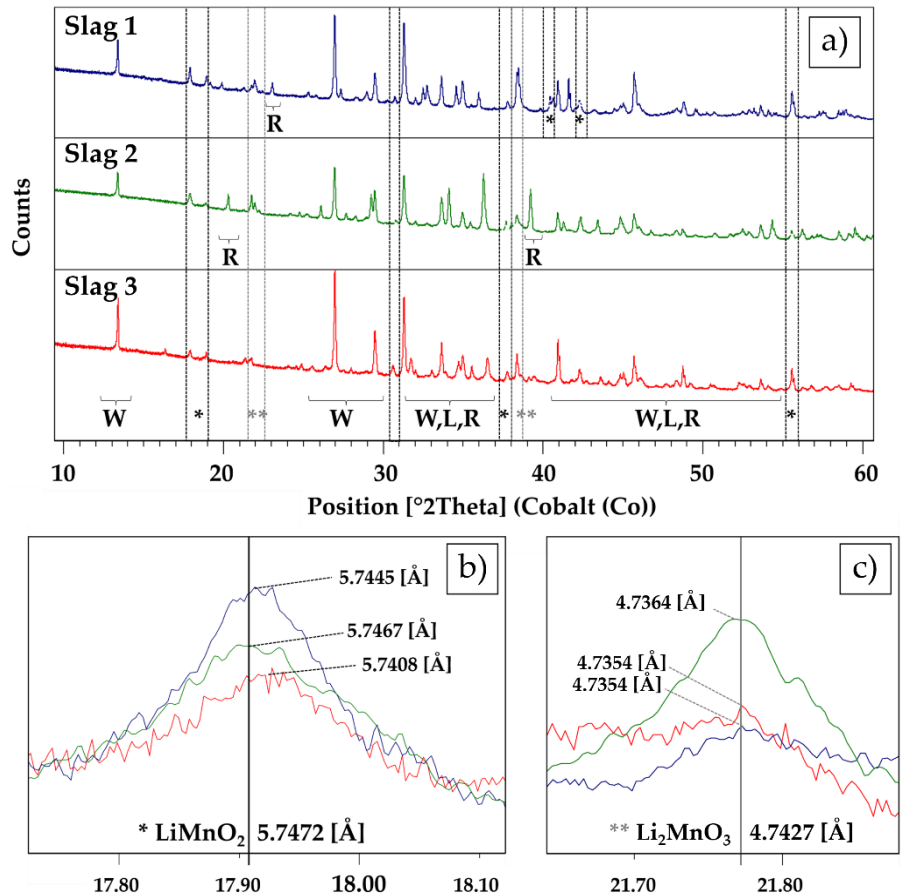

*Figure 5: a) Recorded PXRD pattern of the three slag samples. The identified main phases wollastonite (W), rankinite (R), larnite (L), $LiMnO_2$ (*), $Li_2MnO_3$ (**) were marked. b) Identified $LiMnO_2$ reflex in the slag samples (S1-S3). c) Identified $Li_2MnO_3$ reflex in the slag samples (S1-S3).*

## 3.2 EPMA

With EPMA the following phases were determined:

- $Li_{(1-x)}Mn_{(1+0.33x)}O_2$ / $Li_{(1+x)}Mn_{(1-0.33x)}O_2$ (slag 1,2,3)
- $Li_{(2-x)}Mn_{(1+0.33x)}O_3$ / $Li_{(2+x)}Mn_{(1-0.33x)}O_3$ (slag 2)
- $Li_2SiO_3$ (slag 1,2)
- $Li_xMn_{(3-x)}O_4$ (slag 1,2,3)
- $CaSiO_3$ (slag 1,2,3)
- $Ca_3Si_2O_7$ (slag 1,2)
- $Ca_2SiO_4$ (slag 3)
- Residual melt (slag 2, 3)

These phases were identified based on the elemental content of the individual grains (point measurements, linescans) and the crystal shape. The crystal shape provides additional information regarding the processes during solidification (Figure 6).

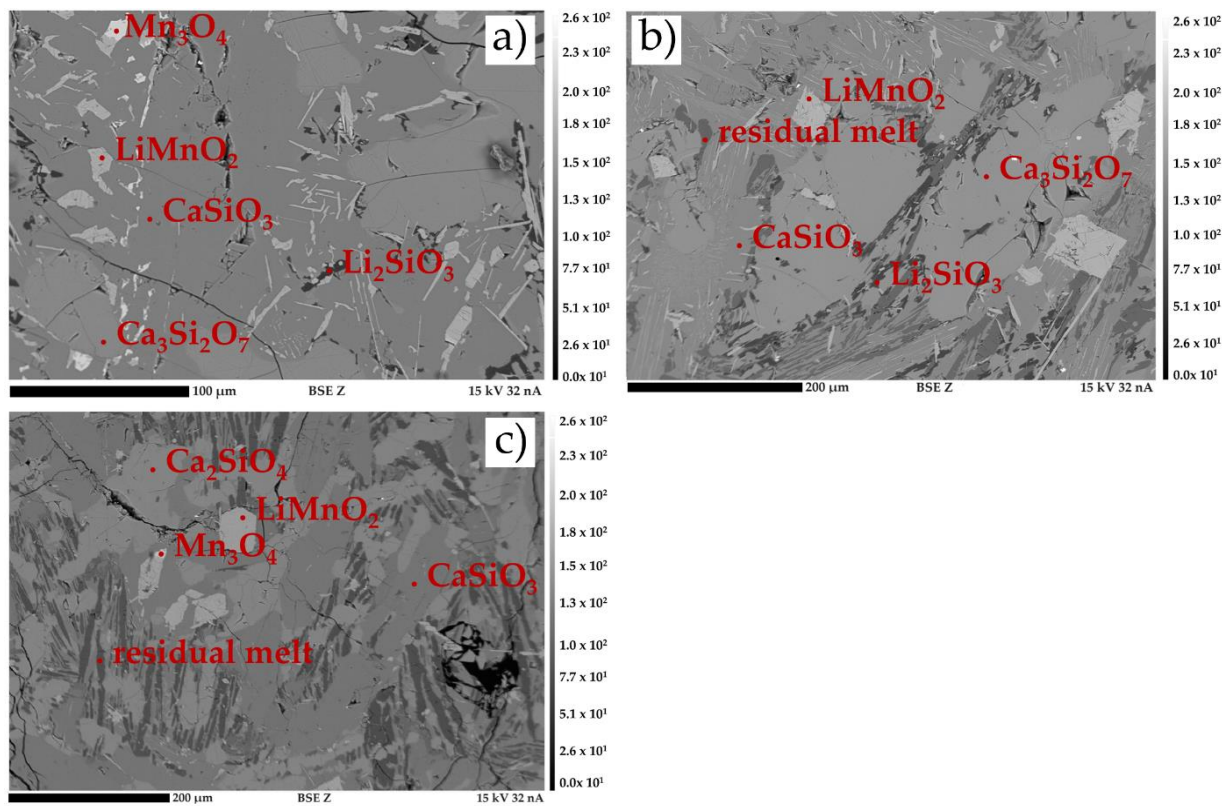

Figure 6: Recorded BSE(Z) micrographs of the three slags. In light gray: hausmannite ($Mn_3O_4$); medium gray: Li-manganate ($LiMnO_2$); dark gray: wollastonite ($CaSiO_3$) and rankinite ($Ca_3Si_2O_7$)/ larnite ($Ca_2SiO_4$). a) S1; b) S2; c) S3.

### 3.2.1 Li-manganates

Two different Li-manganates ($Li_2Mn^{4+}O_3$ and $LiMn^{3+}O_2$) were found in the slag samples. $Li_2MnO_3$ was only detected in slag 2 with EPMA, were 2 wt.% Mg was added. This phase theoretically contains 47.03 wt.% Mn and 11.88 wt.% Li. Within this phase, the average measured Mn content was 47.08 wt.% (min.: 46.88 wt.%; max.: 47.28 wt.%) (Figure 7). In addition, this phase forms idiomorphic crystals between 30 and 100 μm. Based on the crystal shape, it can be determined that $Li_2MnO_3$ forms early during solidification. Point measurements and recorded linescans within this phase show that no other elements (e.g. Mg, Ca, Si) have been incorporated. According to the linescan results, the Mn content within the crystal is constant (Figure 8).

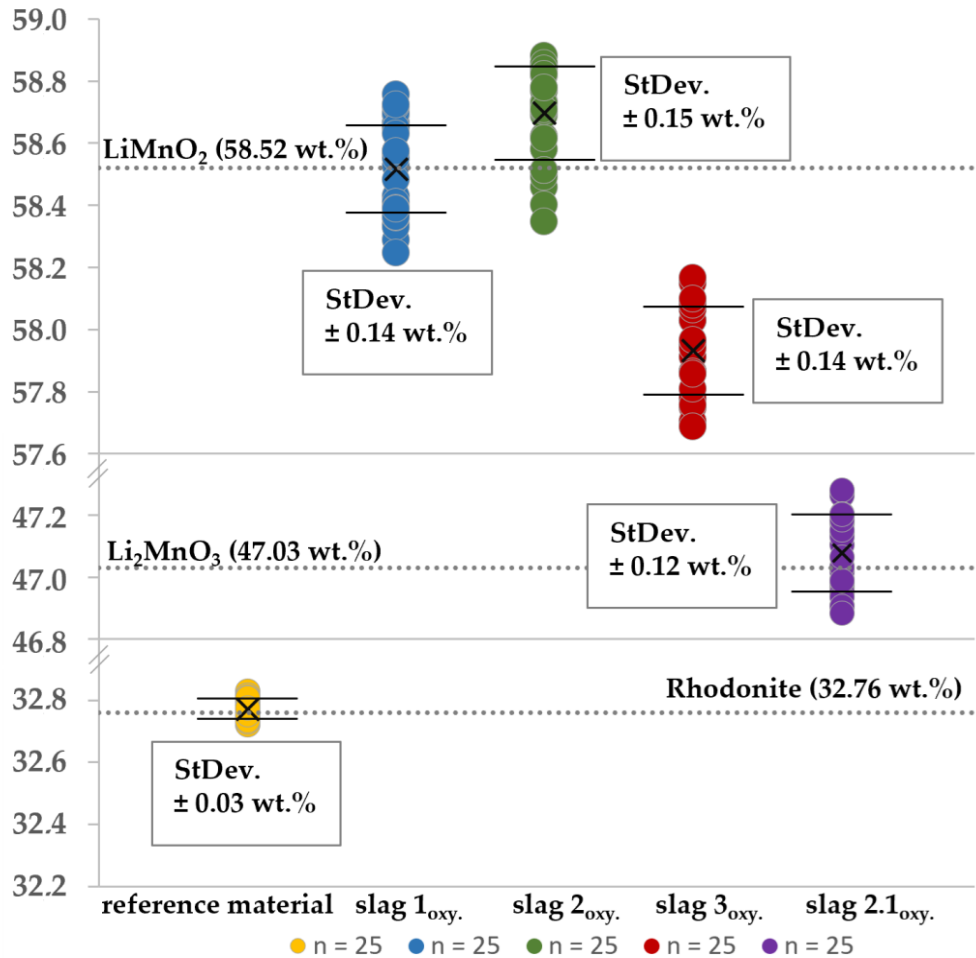

*Figure 7: Overview of the measured Mn concentrations within the Li-manganates in the three slags. The dotted lines indicate the Mn concentration of the pure stoichiometric compounds. In all three slag samples the phase $LiMnO_2$ was found, $Li_2MnO_3$ only in slag 2 (2.1). The shown results based on point measurements in different crystals of the respective phase. Measurements on the reference material rhodonite demonstrate that the Mn content can be measured with an accuracy of ± 0.03 %. X marks the mean value.*

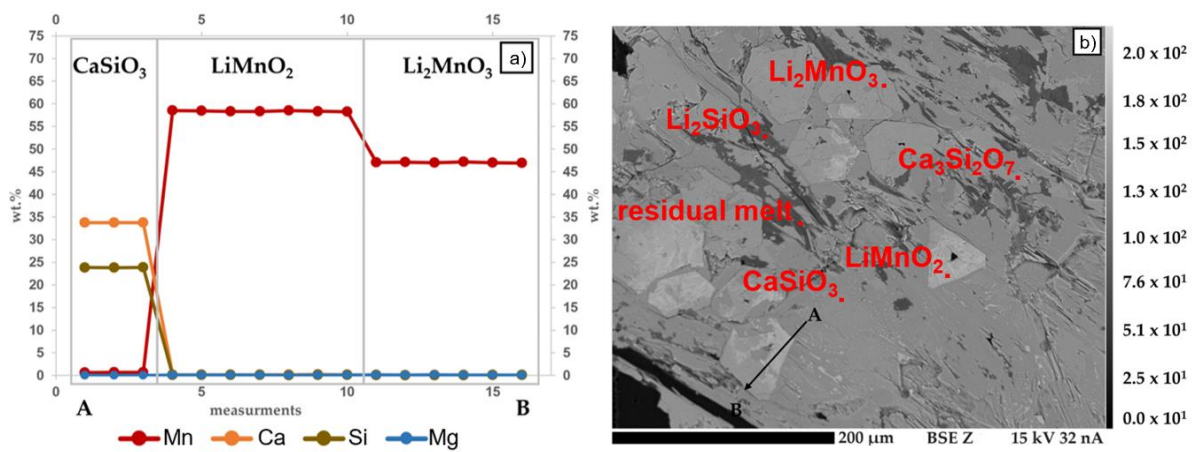

*Figure 8: Elemental composition of the matrix ($CaSiO_3$) and the two Li-manganates ($Li_2MnO_3$) and ($LiMnO_2$) in slag 2. a) The linescan results shown that minimal amounts of $Mn^{2+}$ were incorporated into the matrix. Furthermore, there is no decrease or increase in the element distribution towards the grain boundaries. b) BSE(Z) image showing the path of the line scan (A: start, B: end, step size: 5.3 µm).*

The crystallized LiMnO$_2$ formed idiomorphic to partially hypidiomorphic crystals with an average size of 50 μm. In addition, elongated LiMnO$_2$ needles crystallized in the Ca-silicate matrix. In theory, this phase contains 58.52 wt.% Mn and 7.39 wt.% Li. Within the slag, the average Mn content in S1 was 58.52 wt.% (min.: 58.25 wt.%; max.: 58.76 wt.%), in S2 58.67 wt.% (min.: 58.35 wt.%; max.: 58.88 wt.%) and in S3 57.88 wt.% (min.: 57.02 wt.%; max.: 58.17 wt.%) (Figure 8). Slag 3 contains 0.83 wt.% Al, which results in a lower Mn concentration (Table 3). The results of the linescans shown that the Mn content within the individual grains is almost homogeneous. In addition, no decrease in element concentration is observed towards the grain boundaries (Figure 9).

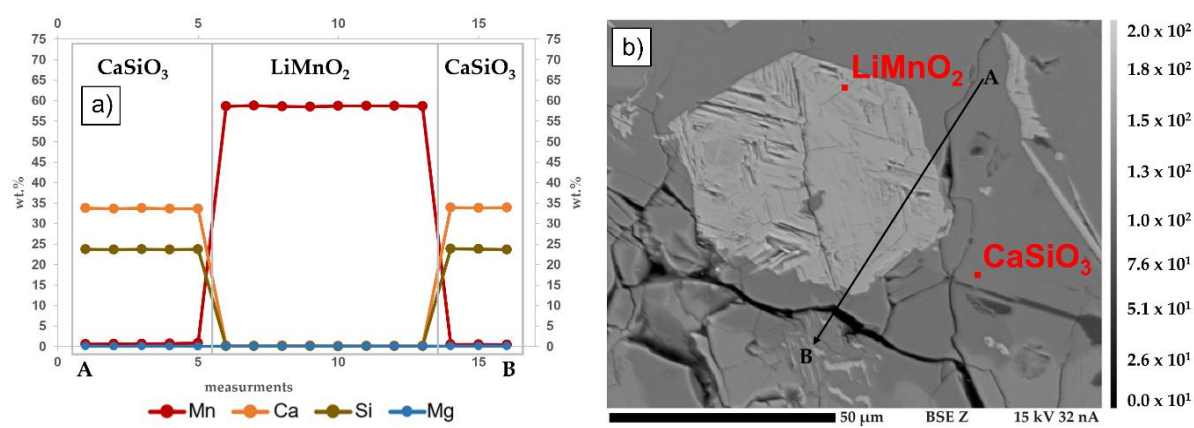

*Figure 9: Elemental composition of the matrix (CaSiO$_3$) and Li-manganate(III) (LiMnO$_2$) in slag 2. a) The linescan results shown that minimal amounts of Mn$^{2+}$ were incorporated into the matrix. Furthermore, there is no decrease or increase in the element distribution towards the grain boundaries. b) BSE(Z) image showing the path of the line scan (A: start, B: end, step size: 4.4 μm).*

*Table 3: For the two Li-manganates LiMnO$_2$ and Li$_2$MnO$_3$, the average structural formula was calculated from all measuring points. To clarify the possible exchange of Li and Mn ions in the crystal lattice, a structural formula was also calculated for the lowest and highest Mn concentration according to: Li$_{(1-x)}$Mn$_{(1+0.33x)}$O$_2$ / Li$_{(1+x)}$Mn$_{(1-0.33x)}$O$_2$ and Li$_{(2-x)}$Mn$_{(1+0.33x)}$O$_3$ / Li$_{(2+x)}$Mn$_{(1-0.33x)}$O$_3$.*

| Sample | LiMnO$_2$ (average) | LiMnO$_2$ (min.) | LiMnO$_2$ (max.) |
|---|---|---|---|
| S1 (n = 25) | Li$^{1+}_{1.0}$Mn$^{3+}_{1.0}$O$_2$ | Li$^{1+}_{1.01}$Mn$^{3+}_{0.99}$O$_2$ | Li$^{1+}_{0.98}$Mn$^{3+}_{1.01}$O$_2$ |
| S2 (n = 25) | Li$^{1+}_{1.0}$Mn$^{3+}_{1.0}$O$_2$ | Li$^{1+}_{1.01}$Mn$^{3+}_{1.00}$O$_2$ | Li$^{1+}_{0.97}$Mn$^{3+}_{1.01}$O$_2$ |
| S3 (n = 25) | Li$^{1+}_{1.0}$(Mn$^{3+}_{0.97}$Al$^{3+}_{0.03}$)O$_2$ | Li$^{1+}_{1.02}$(Mn$^{3+}_{0.96}$Al$^{3+}_{0.03}$)O$_2$ | Li$^{1+}_{0.96}$(Mn$^{3+}_{0.99}$Al$^{3+}_{0.02}$)O$_2$ |
| Sample | Li$_2$MnO$_3$ (average) | Li$_2$MnO$_3$ (min.) | Li$_2$MnO$_3$ (max.) |
| S2 (n = 25) | Li$^{1+}_{2.0}$Mn$^{3+}_{1.00}$O$_3$ | Li$^{1+}_{2.02}$Mn$^{3+}_{0.99}$O$_3$ | Li$^{1+}_{1.97}$Mn$^{3+}_{1.01}$O$_2$ |

### 3.2.2 Hausmannite and spinel

Lithium-rich hausmannite was found in all three slag samples. Hausmannite consists of a mixture of Mn$^{2+}$ and Mn$^{3+}$ occupying tetrahedral and octahedral sites, respectively. When Mn$^{+2}$ is substituted by

Li$^{+1}$, Mn$^{+4}$ can appear in the octahedral sublattice to balance the charge without change of oxygen stoichiometry. However, there is not enough crystallographic data for site occupancies of delithiated hausmannite. Additionally, tetragonal Mn$_3$O$_4$ and orthorhombic LiMnO$_2$ are intergrown together. With an average crystal size of 10-30 µm, these crystals are significantly smaller than LiMnO$_2$. In all three slag samples, the incorporation of other elements into the crystal lattice was detected. The incorporation of Li was recorded in all slag samples. The largest amount of Li was incorporated into S1 with 1.59 wt.%, followed by S2 with 0.80 wt.% and S3 with 0.51 wt.%. In addition, an average incorporation of 0.84 wt.% Mg and 0.82 wt.% Al was observed in S3. For the calculation of the spinel solid solution in S3 the following virtual compounds were defined: Mn$_3$O$_4$, MnAl$_2$O$_4$, MgAl$_2$O$_4$ and Li$_2$Mn$_2$O$_4$. This solid solution can be expressed by the following generalized stoichiometric formula: (Li$_{(2x)}$,Mg$_{(x)}$,Mn$_{(1-2x)}$)$_{1+x}$(Al$_{(2-z)}$,Mn$^{3+}_{(z)}$)$_2$O$_4$ and for Li-rich hausmannite in S1 and S2 these following generalized stoichiometric formula: (Li$_{(2x)}$Mn$_{((1-x))}$)$_{1+x}$(Mn$^{3+}$)$_2$O$_4$ was applied (Table 4).

*Table 4: Determined structural formula of lithiated hausmannite slag in 1,2 and 3.*

| Sample | Mn$^{2+}$Mn$^{3+}_2$O$_4$ |
|---|---|
| S1 (n = 12) | (Li$^{1+}_{0.50}$Mn$^{2+}_{0.75}$)Mn$^{3+}_{2.0}$O$_4$ |
| S2 (n = 16) | (Li$^{1+}_{0.26}$Mn$^{2+}_{0.87}$)Mn$^{3+}_{2.0}$O$_4$ |
| S3 (n = 16) | (Li$^{1+}_{0.16}$Mn$^{2+}_{0.84}$Mg$^{2+}_{0.08}$)(Mn$^{3+}_{1.93}$Al$^{3+}_{0.07}$)O$_4$ |

### 3.3.3 Li-silicate

The Li-silicate Li$_2$SiO$_3$ was only found in S1 and S2. This phase forms 20 to 50 µm rounded, elongated to needle-shaped crystals. The hypidiomorphic to xenomorphic crystal shape indicate crystallization towards the end of solidification. Ideal crystals of this phase contain an average Si content of 31.22 wt.% and an average Li content of 15.43 wt.%. In S1 the average Si content is 31.66 wt.% and in S2 31.68 wt.%. A higher Si content indicate a lower Li content.

### 3.3.4 Matrix forming phases

The main matrix forming mineral is wollastonite (CaSiO$_3$) followed by rankinite (Ca$_3$Si$_2$O$_7$) in S1 and S2 and larnite (Ca$_2$SiO$_4$) in S3. All three phases form xenomorphic crystals, which indicates crystallization at the end of solidification. Due to approximately equal ionic radii of Mg$^{2+}$, Ca$^{2+}$ and Mn$^{2+}$, Ca can be replaced by Mg and Mn in the crystal lattice [35]. In S1 0.85 wt.% Mn was incorporated in wollastonite and 1.28 wt.% in rankinite, in S2 0.59 wt.% Mn in wollastonite and 2.22 wt.% in rankinite and in S3 0.75 wt.% Mn in wollastonite and 6.97 wt.% Mn and 0.76 wt.% Mg in larnite (Table 5).

*Table 5: Calculated structural formulas for the matrix-forming phases wollastonite ($CaSiO_3$) and rankinite ($Ca_3Si_2O_7$) in S1 and S2 and of larnite ($Ca_2SiO_4$) in S3.*

| Sample | $CaSiO_3$ |
| --- | --- |
| S1 (n = 40) | $(Ca^{2+}_{0.99}Mn^{2+}_{0.02})Si^{4+}_{0.99}O_3$ |
| S2 (n = 35) | $(Ca^{2+}_{0.99}Mn^{2+}_{0.01})Si^{4+}_{1.0}O_3$ |
| S3 (n = 30) | $(Ca^{2+}_{0.99}Mn^{2+}_{0.02})Si^{4+}_{1.0}O_3$ |
| Sample | $Ca_3Si_2O_7$ |
| S1 (n = 40) | $(Ca^{2+}_{2.96}Mn^{2+}_{0.07})Si^{4+}_{1.99}O_7$ |
| S2 (n = 40) | $(Ca^{2+}_{2.86}Mn^{2+}_{0.12}Mg^{2+}_{0.02})Si^{4+}_{2.0}O_7$ |
| Sample | $Ca_2SiO_4$ |
| S3 (n = 28) | $(Li^{1+}_{0.33}Ca^{2+}_{1.63}Mn^{2+}_{0.14}Mg^{2+}_{0.05})Si^{4+}_{0.99}O_4$ |

## 4. Discussion

The aim of this study was to investigate the stabilization of $Mn^{4+}$ in the slag and to verify also which Li-rich phases are formed. Moreover, it should be investigated whether $Mn^{4+}$ forms a compound with Mg or Al. Based on the phase diagrams, the experiment provides an overview of which phases crystallize in a synthetic slag consisting of $Li_2O$, $MgO$, $Al_2O_3$, $SiO_2$, $CaO$ and $MnO_x$.

### 4.1 Influence of Mg and Al on $Mn^{4+}$

As can be seen from Figure 3e, the spinel phase in the $MgO$-$Al_2O_3$ system is stable phase from low temperatures and has an extension towards both $Al_2O_3$ and MgO concentrations at high temperatures. The addition of MgO together with $Al_2O_3$ will increase stability range of cubic spinel in the $Li_2O$-$MnO_x$-$MgO$-$Al_2O_3$ system.

The tetragonal-to-cubic transformations for the spinel phase in the $MgO$-$MnO_x$ and $Al_2O_3$-$MnO_x$ systems are driven by the Jahn-Teller distortion of octahedral sites, primarily occupied by $Mn^{3+}$ ions. In the cubic phase, assuming a disproportionation of $Mn^{3+}$ into $Mn^{4+}$ and $Mn^{2+}$ and occupation the octahedral sites by $Mn^{+4}$ and distribution of $Mn^{+2}$ between tetrahedral and octahedral sites, a general formula for cubic spinel $(Al^{3+}, Mn^{2+})_1 (Al^{3+}, Mn^{3+}, Mn^{2+}, Mn^{4+}, Va)_2 (Mn^{2+}, Va)_2 (O^{2-})_4$ is employed in

$Al_2O_3$-$MnO_x$ system and $(Mg^{2+}, Mn^{2+})_1 (Mg^{2+}, Mn^{3+}, Mn^{2+}, Mn^{4+}, Va)_2 (Mg^{2+}, Mn^{2+}, Va)_2 (O^{2-})_4$ in MgO-$MnO_x$ system. This formulation accommodates the variation in oxidation states of Mn within the crystal lattice. The cubic spinel phase is further characterized by the inclusion of vacancies (Va) on the octahedral sites and $Mn^{+2}/Mg^{+2}$ in interstitial site. The vacancies are introduced into the spinel model to extend the homogeneity ranges towards $Al_2O_3$ side and interstitial site to model extension towards $MnO_x$/MgO side. The applied models account for deviations from a perfectly ordered structure to describe disordering in cationic sites and to describe homogeneity ranges of spinel thus enhancing the model applicability in a large range of conditions (compositions, temperature, oxygen partial pressure). Modelling of tetragonal spinel differs from cubic by introducing $Mn^{+3}$ into tetrahedral site and by absence of $Mn^{+4}$ in octahedral site in both systems. Tetragonal phase is stable at lower temperatures and in more narrow composition range than cubic spinel.

The octahedral sites are substantially occupied by $Mn^{4+}$ in cubic spinel in the MgO-$MnO_x$ system in the whole possible composition range. Low concentration of $Mn^{4+}$ is found in the spinel phase with compositions close to stoichiometric $MnAl_2O_4$ in the $Al_2O_3$-$MnO_x$ system even at high oxygen partial pressure. Thus, the introduction of Mg into the spinel phase not only enhances the stability but also effectively increase the $Mn^{4+}$ concentration. This suggests a more pronounced role of Mg in maintaining the desired Mn oxidation state within the spinel structure.

The Li addition introduces a notable impact on the $Mn^{4+}$ ion concentration on the octahedral sites. Substitution of $Mn^{+2}$ for $Li^{1+}$ in the tetrahedral sites reduces the charge of cations in spinel. In order to preserve the electroneutrality of this structure the fraction of $Mn^{+4}$ in octahedral sites occupied by $Mn^{+3}$ in tetragonal spinel should be increased. The increase of $Mn^{+4}$ in octahedral site should also reduce Jahn-Teller effect caused by electronic structure of $Mn^{+3}$ cation. It should be noted that in the stoichiometric spinel $LiMn_2O_4$ cation $Li^{+1}$ completely occupies tetrahedral site and the ratio of $Mn^{+4}/Mn^{+3}$ in octahedral site is equal to one. In case of Li content is higher than in stoichiometric spinel, $Li^{+1}$ partially occupies octahedral site and ratio $Mn^{+4}/Mn^{+3}$ becomes larger than one. With the temperature increase spinel becomes enriched by Mn with the shift of composition to stoichiometric spinel $LiMn_2O_4$ and even higher content of Mn. Therefore, spinel with high Li content stable at lower temperature, with the temperature increase will decompose to spinel with lower Li content and $Li_2MnO_3$ phase (Figure 2). At temperature 1230 K $LiMnO_2$ forms from spinel and $Li_2MnO_3$ and at temperatures 1250 K, $Li_2MnO_3$ decomposes into $LiMnO_2$ and $Li_2O$. Tetragonal spinel with composition close to stoichiometric one is stable in equilibrium with hausmannite with much lower Li content from one side and from other side it is in equilibrium with $LiMnO_2$. High temperature stability limit of tetragonal spinel is defined by decomposition to hausmannite and $LiMnO_2$ at 1270 K.

According to the phase diagram shown in Figure 3b, no reactions should take place between $MnO_2$ and $Al_2O_3$ up to 688 K. The experimental results indicate that Mg and Al react to form spinel ($Mg_2AlO_4$) and therefore small amounts of these elements are not a problem for Li-manganate ($LiMnO_2$ or $Li_2MnO_3$) forming processes. As long as $Mn^{2+}$ and $Mn^{3+}$ are present in the slag, spinels with Mg, Al and Mn are formed, which can interfere with the formation of EnAMs. In order to produce large and pure EnAM crystals that can also be efficiently separated from the rest of the slag, it must be ensured that $LiMnO_2$ or $Li_2MnO_3$ crystallize out of the melt first.

### 4.2 Li-rich phases as a new potential EnAM

A total of four Li-containing phases crystallized out in the slag. These phases include the two Li-manganates $LiMnO_2$ and $Li_2MnO_3$, as well as the Li-silicate $Li_2SiO_3$ and Li-containing hausmannite. With an average of 0.51 wt.% (S3) to 1.59 wt.% Li (S1), the Li content in hausmannite is too low to be a potential EnAM. The $Li^{+1}$ is incorporated at the tetrahedral site and replaces $Mn^{2+}$ in the crystal lattice. The incorporation of $Li^{+1}$ into cubic $LiMn_2O_4$ can occur by balancing $Mn^{4+}$ on the octahedral site, but it also can occur by deviation from stoichiometry of oxygen by introducing vacancies [36–38]. There is not enough crystallographic data for hausmannite showing how substitution of $Mn^{+2}$ by $Li^{+1}$ is compensated. Since it is not possible to detect small amounts of $Mn^{4+}$ with EPMA, more advance technique should be used for further investigated whether $Mn^{4+}$ can be incorporated into the crystal lattice of hausmannite for charge compensation. Hausmannite forms in addition relatively tiny crystals and only occurs in small amounts. Due to the partially inhomogeneous mixture of $Mn^{2+}$ and $Mn^{3+}$ in this phase, further processing would be necessary after successful separation from the slag. With an average of 15.43 wt.% Li, $Li_2SiO_3$ would be more suitable as a new EnAM. However, this phase crystallizes towards the end of solidification and thus forms relatively small and needle-shaped crystals, which is an inhibitor for a successful separation of Li. Furthermore, Li and Si must be separated from each other after the recycling step. Therefore, an Li-manganate like $LiMnO_2$ or $Li_2MnO_3$ would be more suitable as new potential EnAMs. The Li-manganate $Li_2MnO_3$, with 11.88 wt.% Li, could only be found in slag 2, where 2 wt.% Mg was added. No incorporation of other elements could be observed within this phase. As shown in Figure 2, the $Li_2MnO_3$ phase is stoichiometric which corroborates with the absence of solubility of other elements. Furthermore, the $Li_2MnO_3$ phase has some advantages as a new EnAM. Firstly, it is an early crystallite with large, idiomorphic crystals and containing more Li compared to $LiAlO_2$ (10.35 wt.% Li). The stabilization of only $Mn^{4+}$ in the slag has the additional advantage that spinel formation with Mn could be suppressed and that the Jahn-Teller effect does not cause deformation. In addition, $LiMnO_2$ (7.39 wt.% Li) can also be considered as a potential EnAM. The low incorporation of Al (0.83 wt.%) does not seem to have any effect on the crystal lattice. According to Kong et al. [16], small amounts of Al in the Li-rich cathode material should promote intralayer diffusion.

Nevertheless, it remains to be determined which of the two Li manganates, $LiMnO_2$ or $Li_2MnO_3$, formed first. It is possible that $LiMnO_2$ forms first as a high temperature phase and remains stable during cooling. At low temperatures, according to phase diagram the stable phases are spinel and $Li_2MnO_3$. The alternative would be that $Li_2MnO_3$ forms first and decomposes to $LiMnO_2$ and $Li_2O$, which reacts with $SiO_2$ forming $Li_2SiO_3$.

## 5. Conclusions and outlook

The presented results indicate that it was possible to stabilize the $Mn^{4+}$ containing Li-manganate $Li_2MnO_3$ in a synthetic slag. However, the conditions under which $Li_2MnO_3$ was formed and how $Mn^{4+}$ was stabilized throughout the slag remain to be clarified. A promising approach would be to investigate phase relations (experimentally and theoretically) and to develop thermodynamic databases for a complex system (e.g. a slag system consisting of several elements such as Li, Mg, Al, Si, Ca, Mn, O).

The interpretation of these data could help to understand the processes taking place during solidification. This could help to optimize the slag system and the cooling curves to maximize the EnAM yield.

Both Li-manganates, $Li_2MnO_3$ and $LiMnO_2$ have proven to be extremely pure phases. Accordingly, the added oxygen only had an influence on the regions close to the surface, where $Li_2MnO_3$ could find. Therefore, the viscosity of the slag should therefore be reduced in ongoing experiments so that the oxygen can influence the entire slag in subsequent experiments. To achieve this, small amounts of FeO and $Fe_2O_3$ will be added to the precursors (e.g. [39]). With the help of the added iron, more oxygen should enter the slag and mainly $Mn^{4+}$ should be stabilized. In addition, more investigations are necessary to verify whether $Mn^{4+}$ forms a spinel-like compound with Al or whether the formation of spinels could be suppressed by oxidizing Mn to $Mn^{4+}$. This includes determining at what wt.% Mg and Al inhibit the formation of Li-manganates. In addition, it must be determined how much $Mn^{4+}$ can be incorporated into the crystal lattice of the spinels and which phases crystallize first on solidification of the slag - spinels between $Mg^{2+}$, $Al^{3+}$, $Mn^{2+}$ and $Mn^{3+}$ (as well as $Mn^{4+}$) or Li-manganates.

Declaration of competing interest

The authors declare that they have no known competing financial interests or personal relationships that could have appeared to influence the work reported in this paper.

Acknowledgements

The authors gratefully acknowledge the financial support from DFG (Deutsche Forschungsgemeinschaft) in frame of the subproject SPP-2315: EnAM (Engineered Artificial Minerals) (ProjNo. 470309740 and 470392360) and the support by Open Access Publishing Fund of Clausthal University of Technology.